\documentclass[11pt,letterpaper]{article}
\usepackage{makecell}
\usepackage[latin1]{inputenc}
\usepackage{amsmath}
\usepackage{amsfonts}
\usepackage{amsthm, amssymb, amscd}
\usepackage{hyperref}
\usepackage{graphicx}
\usepackage{booktabs}
\usepackage[]{algorithm2e}
\hypersetup{colorlinks=true, citecolor=blue, urlcolor=blue,linkcolor=blue}

\newtheorem{lemma}{Lemma}
\newtheorem{definition}{Definition}

\newtheorem{remark}{Remark}

\newcommand{\footremember}[2]{%
    \footnote{#2}
    \newcounter{#1}
    \setcounter{#1}{\value{footnote}}%
}

\author{
	Carenne Lude\~na\footremember{ujtl}{Departamento de Ciencias Básicas. Facultad de Ingeniería y Ciencias Básicas.
		Universidad Jorge Tadeo Lozano, Carrera 4ta \#22-61, Bogotá, Colombia.
	} \footnote{Corresponding author}\,\,,
	Miguel Mendez\footremember{uan}{Facultad de Ciencias, Universidad Antonio Nari\~no, Sede Circunvalar, Cra 3 Este \# 47 A-15, Bogot\'a, Colombia.
	}\,,
	Nicolas Bolivar\footremember{uan1}{Centro de Investigación en Ciencias Básicas y Aplicadas, Universidad Antonio Nari\~no, Sede Circunvalar, Cra 3 Este \# 47 A-15, Bogot\'a, Colombia.}
}

\title{Modular decomposition of graphs and hierarchical modeling}
\begin{document}
\date{}
\maketitle
\begin{abstract}
	We consider Gallai's graph Modular Decomposition theory  for network analytics. On the one hand, by arguing that this is a choice tool for understanding structural and functional similarities among nodes in a network. On the other, by proposing a model for random graphs based on this  decomposition. Our approach establishes a well defined context for  hierarchical modeling and provides a solid theoretical framework for probabilistic and statistical methods. Theoretical and simulation results show the model acknowledges scale free networks, high clustering coefficients and small diameters all of which are observed features in many natural and social networks.
\end{abstract}
\noindent {\bf Keywords:} Complex networks, Modular decomposition, Random graph models, Hierarchical graph models, scale free models.
\section{Introduction}  
Networks have become  ubiquitous in modeling complex structures. Applications range from social networks, to energy or transportation grids, proteomics or genetics, chemistry or brain structure. Networks are popular because they are able to address not only the characteristics of the units or individuals composing the system but also their interactions. In this graph representation, units are considered as nodes and their interactions as edges between these nodes, where the latter can be  non directed or directed. 

Real systems tend to be highly non trivial, characterized by small diameters, high average clustering coefficients and scale free behavior (\cite{estrada,AB2}). Also,  in many applications, ranging from social networks to genomic, proteonomic or brain networks \cite{bnet}, there is a natural hierarchical structure induced by functional, spatial or other types of relationships that is not readily obtained by most popular graph models such as the classical ER random graph models (\cite{ER1}, \cite{estrada}), scale free models  (\cite{AB1}, \cite{AB2}, \cite{estrada}) or exponential random graph models, ERGMs (\cite{egms1}). This has lead to a series of models which explicitly include a hierarchical structure  such as Hierarchical ERGMs (\cite{hergms}), multifractal RGMs (\cite{palla}), k-core based models \cite{karwa} or stochastic Block models (\cite{han},\cite{SBM1}). The latter are not solely interested in the description of single nodes and their connectivity, but rather in the appereance of  subsets of nodes with similar characteristics connected hierarchically. These subsets are assumed latent unobserved classes or blocks.

The idea of considering connections among blocks of nodes has a natural interpretation in classical discrete mathematics. Indeed,  for a variety of structures such as graphs, directed graphs, partially ordered sets, boolean functions, hypergraphs, multipersonal simple games, etc., it is possible under certain conditions to define a decomposition based on substitution of structures in a nested schema:  the complete structure can be decomposed as a simpler outer structure each of whose components represent in turn internal structures (see for example \cite{gallai},  \cite{mh1} and \cite{mr1}). More recently, in \cite{mm1} this decomposition of finite structures has been formulated in the context of Operad Theory, leading to a very general  notion of a unique  decomposition which can then be thought of as a \emph{factorization}. This will be further discussed in Section 2. In the case of graphs,  it is called the modular decomposition (MD). Gallai, \cite{gallai}, was the first to show that the MD of an undirected, simple graph always exists and is unique.

In a sense, this factorization can be thought of as a generalization of the decomposition of a graph in connected components, including the possibility of modeling other homogeneous relationships such as two-mode or affiliation networks, where connections among the elements of one mode are based on their linkages established through the second mode \cite{wasfaust}. This is interesting because much of classical graph models study mostly modular type behavior, defined by dense relationships among vertex within modules and sparce relations among modules, but this leaves out many  functional-like relationships. 

Modular decomposition has been used for graph applications such as graph drawing \cite{graphdrawing} or calculating distances between graphs \cite{largegraphs}.  It has also been used to identify logical relations among the members of the modules, e.g. exclusive alternatives for the rest of the network \cite{gagneur}, as well as interesting structures   such as cliques \cite{size}. Finally, the MD of a graph is known to be achievable in linear-time \cite{habib,tedder1}, which is desirable for large-scale networks, as those encountered in social or natural networks such as the brain or in proteomics.  

As mentioned, the MD of a graph has been used in a series of applications. However,  it doesn't seem to have entered the mainstream of graph analytics or modeling. The  objective of this article is then twofold. On the one hand, introducing MD as a choice hierarchical tool for understanding structural and functional similarities among nodes, generalizing previous efforts in this direction. On the other,  to propose a model for random graphs based on this factorization. Our approach establishes a well defined context for  hierarchical modeling and provides a solid theoretical framework for probabilistic and statistical methods. 

The article is organized as follows. In Section \ref{MD} we introduce the basic theoretical background for MD.  In section \ref{exps} we present statistics for  the MD of simulated and real graphs. In Section \ref{secmdrg} we define the model and discuss some of its properties. Finally in Section \ref{simul} we present simulations with the model introduced un Section \ref{secmdrg}. In Section \ref{conc} we present  concluding remarks.

\section{Modular decomposition of graphs}\label{MD}

A discrete structure  \emph{factorization} is  in each case associated with a product that yields a hierarchy of nested structures that can be represented as a rooted tree. In the case of undirected graphs, the factorization is associated with the product described as follows. Denote by $V$ the set of vertices and consider a set partition $\pi$ of $V$ and a graph $g_B$ on each block $B$ in $\pi$. 

\begin{figure}[h]
	\begin{center}
		\includegraphics[width=12cm]{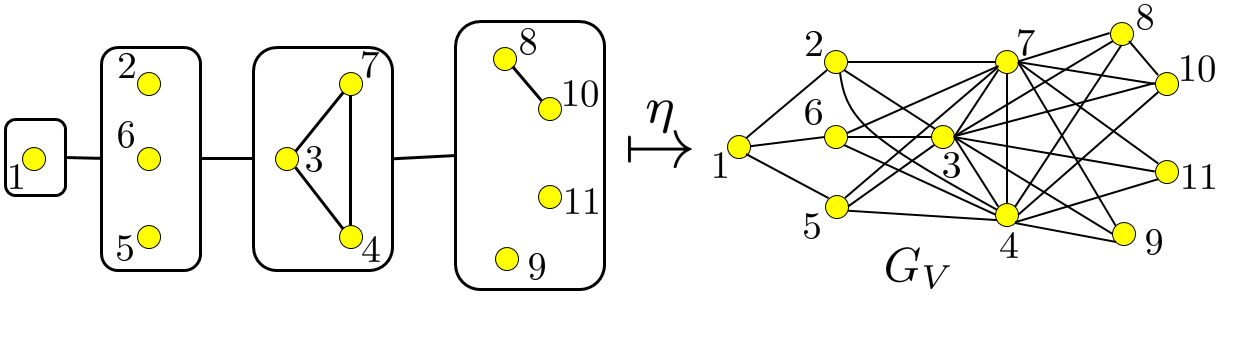}\end{center}\caption{The product of nested graphs, $V=\{1,2,\dots,11\}.$}\label{Fig.factorization1}
\end{figure}

Consider also a graph $G_{\pi}$ with `fat' vertices; its vertices are the blocks of $\pi$. The whole structure has the form $$(\{g_B\}_{B\in\pi}, G_{\pi}).$$ It is a nested graph, thought of as an \emph {external} graph $G_{\pi}$  whose vertices are the blocks of the partition, each block provided itself with a graph (we will call them internal or inner graphs, see Fig.\ref{Fig.factorization1}). Then we define the product \begin{equation}\label{Eq.factorization}(\{g_B\}_{B\in\pi}, G_{\pi})\stackrel{\eta}{\mapsto} G_V\end{equation}\noindent as the graph obtained by keeping all the edges of the inner graphs plus some more extra edges created by using the external graph $G_{\pi}$. For each external edge $\{B, B'\}$ of $G_{\pi}$ add all the edges of the form $\{b,b'\}$, $b\in B$ and $b'\in B'$ (see Fig. \ref{Fig.factorization1}). Each block of $\pi$ is a \emph{ module} of $G_V$ in the sense of the following definition.

	\begin{definition}Let $G_V=(V,E)$ be a graph with vertices in $V$. A module of $G_V$ is a subset of vertices $M\subseteq V$ that satisfies for each $v\in V-M$  either $v$ is connected to all $b\in M$  or is not connected to any. 
	\end{definition}
The singletons subsets and the whole set $V$ are modules of $G_V$. They are called the trivial modules.  A non-trivial module $M$ gives rise to a non-trivial factorization of the graph $G_V$ by taking the partition with only one big block $M$, the rest of them singletons.  Now we define  prime graphs.
	\begin{definition}
		A graph $G_V$ is called prime if it does not have modules other than the trivial ones.
	\end{definition}

	It is easy to check that if $G_V$ is prime, the only possible factorization, Eq. (\ref{Eq.factorization}) are the trivial ones. Either $\pi$ is the partition of singletons,  $\pi=\{\{v\}|v\in V\}$ the internal graphs  being the trivial ones and $G_{\pi}=G_V$, or $\pi=\{V\}$, the internal graph $G_V$ and $G_{\pi}$ the trivial graph with only one `fat' vertex. The reader may check that the path $P_4$ is prime, the path $P_3$ is not.

\begin{figure}[h]
	\begin{center}
		\includegraphics[width=12cm]{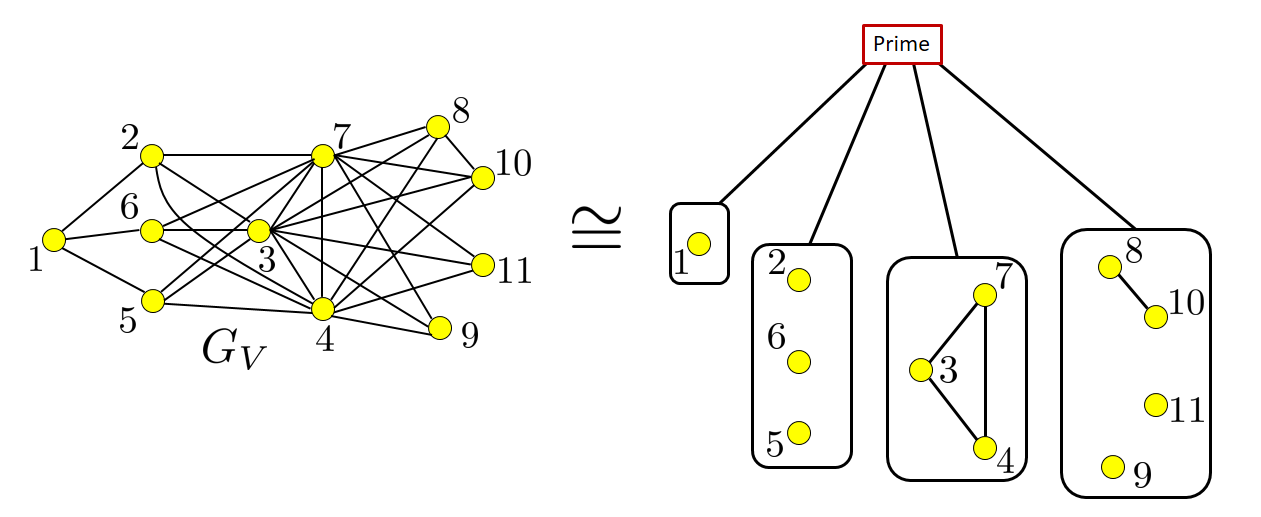}\end{center}\caption{Small treee.}\label{Fig.factorization2}
\end{figure}

\begin{figure}[h]
	\begin{center}
		\includegraphics[width=12cm]{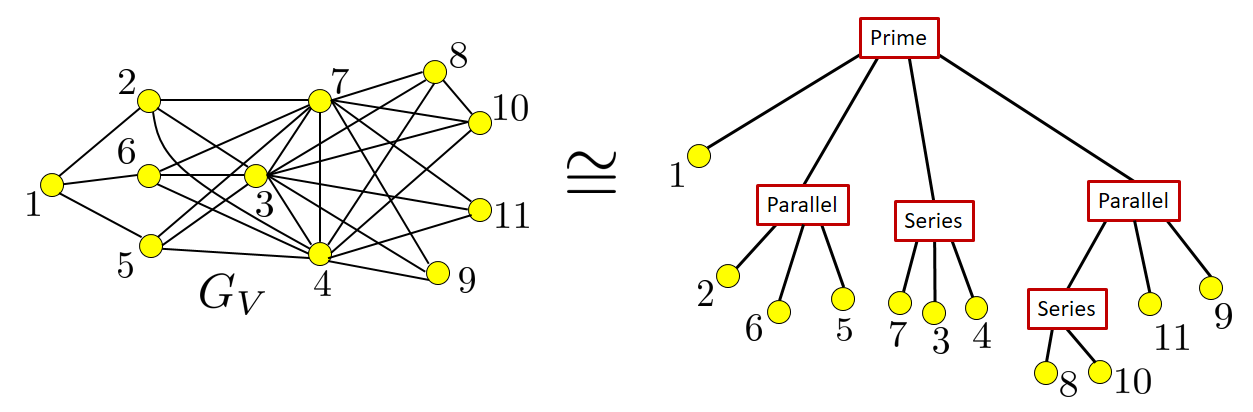}\end{center}\caption{Modular decomposition tree (MD treee) of $G_V$.}\label{Fig.factorization3}
\end{figure}

Looking at the product map $\eta$ in Fig \ref{Fig.factorization1}, we get a factorization of the graph $G_V$ that can be represented as a small tree whose root has the tag `prime', since the graph $P_4$ is prime (in general, the information of the specific prime graph is kept together with the tag `prime'). The inner graphs are its children  (see Fig. \ref{Fig.factorization2}). If the outer graph is complete we tag the root vertex `series' and `parallel' if it is an empty (edgeless) graph. The procedure of factorization can be performed iteratively on each of the inner graphs until we reach singletons, wich become  the leaves of a tree (see Fig. \ref{Fig.factorization3}). 

Gallai \cite{gallai} showed that this factorization is unique if 
\begin{enumerate}\item We use only modules of three tag types: \emph{prime}, \emph{series} (a complete graph), \emph{parallel} (edgeless, empty graph).
	\item Neither a complete graph is a child of a complete graph, nor an empty graph is a child of an empty graph.
\end{enumerate}  
A decomposition of a graph in a tree as described above will be called a {\em modular decomposition} (MD).

The notion of a module and the idea of modular decomposition for graphs have been rediscovered many times
in different settings starting from the seminal work of Gallai \cite{gallai} and considering generalizations to other discrete structures (see for example \cite{mr1,mh1}).   In the  context of Operad Theory and combinatorial species, Mendez \cite{mm1}, discusses general decomposition theory including graphs and many other discrete structures such as digraphs,  boolean functions, relational structures and set systems.
  
This general setting moreover  provides a relationship between the generating functions of prime graphs and general simple graphs over $n$ labeled vertices, which we omit being out of the scope of this paper. This allows to count the actual number of prime graphs for a given  number of vertices. These values are given in Table \ref{tablaprimos} up to  $m=15$ (starting from $m=4$ as there are no prime graphs for $m<4$).  Proportions are calculated using the number of graphs with n vertices is $2^{\binom{n}{2}}$. As can be seen, the proportion of prime graphs tends to one quite rapidly, so that prime graphs are dominant over the set of graphs. Two interesting conclusions can be deduced. The first, is that a good way to sample prime graphs is to consider a uniform distribution over the set of graphs  and then use an acceptance-rejection procedure, accepting the simulated graph only if it is prime. Since the uniform distribution over the set of all graphs is obtained by using the ER random graph model \cite{ER1}, with $p=0.5$ this provides a simple and asymptotically quite efficient algorithm which will be discussed in detail in Section  4.

 The second,  is that since most natural or social graphs tend to have  complex structures with several layer MD trees, uniform probabilities over the set of graphs are quite unrealistic (this fact had been previously signaled by M\"ohring, pg. 5 \cite{mr1}). As we shall see in Section \ref{exps}, for large size connected graphs,  the most common structure is a MD tree with a prime root followed by more simple parallel or series nodes. Here primes can be very different, and  the form of this prime root is  fundamental  to understand the overall graph structure. Moreover, changing the random graph model in simulations has an important impact on the structure of the prime root in connected graphs.
 
 Subsequent nodes in the MD tree are also important in terms of understanding graph structure and can be interpreted in biological or social terms (see \cite{gagneur} for an application to a protein interaction network). A series  node can be thought of as a single  unit in the structure with all sub-units completely connected. A parallel node can be interpreted as alternative sub-units, any of which plays the same role in the structure. In a sense, this can be thought of as an \emph{information replication}  strategy, where certain selected connections in the outer structure  are repeated by non-interacting sub-units.
 
\begin{table}
	\begin{center}
		\begin{tabular}{|c|c|c|}
			\hline \hline
			$m$ & \# of primes & proportion of primes\\
			 4 & 12 & 0.1875\\ \hline 
			 5 & 192 & 0.1875\\ \hline 
			6 & 10800 & 0.3295\\ \hline 
			 7 & 970080 & 0.4625\\ \hline 
			 8 & 161310240 & 0.6009\\ \hline 
			10 & $28*10^{12}$ & 0.8153\\ \hline
			15 & $39*10^{30}$ & 0.9855 \\		\hline 
		\end{tabular}
	\end{center}
	\caption{Number and proportion of labeled \emph{prime} graphs.}
	\label{tablaprimos}
\end{table}

\section{Modular decomposition examples}\label{exps}

In order to understand how the MD is related to graph structure, we considered a series of examples of both real and simulated graphs. The MD algorithm we used is based on \cite{mdalg}. Code was developed in java and used in an R script. The algorithm provides the MD tree and the following statistics: number of primes, series and parallel nodes in the decomposition, tree depth and length of largest prime node. Statistics are calculated for $N=50$ simulations. Mean and s.d. values are presented. Graphical representations of the MD trees were constructed using packages Igraph, data.tree and networkD3 in R.

\subsection{Real graphs}

As mentioned, the literature of MD for graph analysis is scarce. A notable exception  is given in \cite{gagneur}. The authors in this article apply MD to help understand the  structure within and between detected regions of  the   protein interaction network displaying
221 interactions involving 131 proteins of the
human TNF-$\alpha$/NFkB signal transduction pathway. Their findings include a prime root (with 5 children), followed by parallel and series structures. A thorough biological based analysis of functional equivalence of studied units suggests that series modules can be thought of as functional single subunits, whereas parallel modules can be interpreted as alternative structures (functional equivalence) \cite{gagneur}, pg. 10.

We consider four examples of natural networks and the Zachary Karate club graph \cite{zac} as a (small) example of a social network. The first natural network, corresponds to the transcription factor-gene interaction network of \emph{Escherichia coli} from \hspace{20pt}
{\small \verb|http://regulondb.ccg.unam.mx/menu/download| }
\break	 {\small \verb|/datasets/index.jsp|}
 (\cite{regulonDB}). The nodes of the network are Transcription Factors (TF), understanding that TF can regulate their own transcription. Edges indicate the existence of regulation, where the regulatory effect of the TF can be  (+) activator, (-) repressor, (+-) dual or (?) unknown.  All loops are eliminated.

 \noindent The other three, are the  mouse visual cortex 1, mouse visual cortex and macaque Rhesus brain 1 network, 3 networks from \verb|https://neurodata.io| \verb|/project/connectomes|. Characteristics of the  first two examples are detailed in \cite{repository}  and the last example is described in   \cite{ros}. 

For all networks the MD tree was obtained as well as general statistics related to the MD tree such as prime, series and parallel densities, number of levels and length of longest prime as a proxy of prime complexity.

\begin{figure}
 	\begin{tabular}{ccc}
 		\includegraphics[width=4cm,height=5cm]{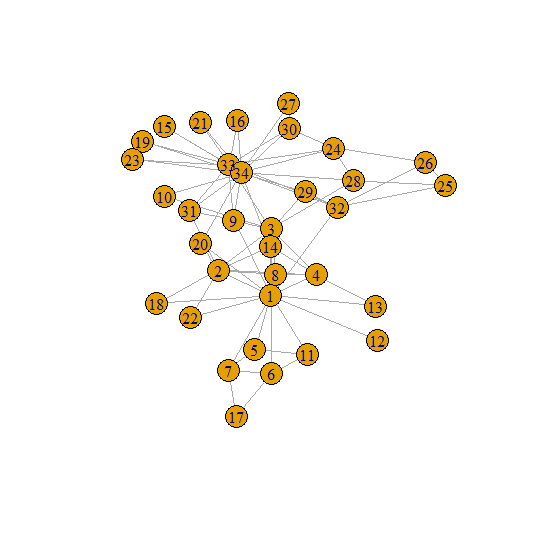}
 		&
 		\includegraphics[width=4cm,height=5cm]{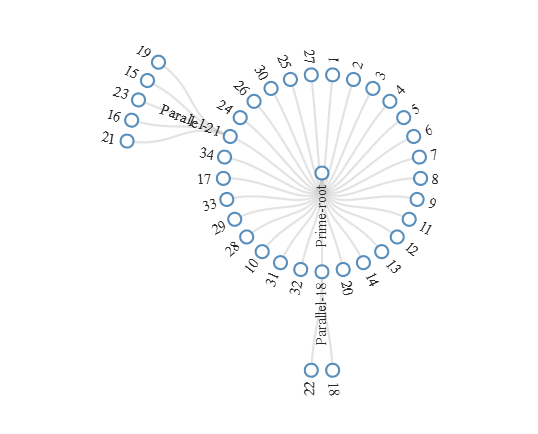}
 		&
 		\includegraphics[width=4cm,height=5cm]{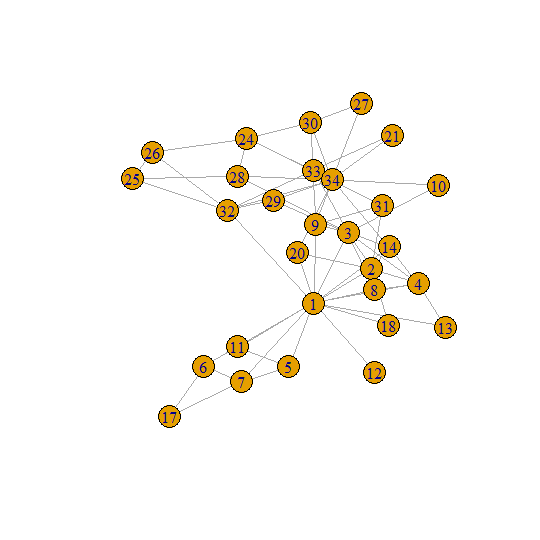}
 		
 	\end{tabular}
 	\caption{ Zachary karate club original graph (left), MD tree for the Zachary graph (center), Largest prime graph at level 1 (right)}
 	\label{figzachary}
 \end{figure}

The Zachary graph along with its MD tree and the graph of its largest prime are shown in Figure \ref{figzachary}. Its MD tree has 3 internal nodes and 2 levels. The root of the tree is \emph{prime}, followed by two \emph{parallel} nodes. The \emph{prime} has 29 nodes (out of 34). And the \emph{parallel} nodes have, respectively 5 children (nodes 15,16,19,21 and 23 in the original graph) and 2 children (nodes 18 and 22 in the original graph). The size 5 node connects to nodes 33 and 34 and the size 2 node connects to nodes 1 and 2 in the original graph. All members of the \emph{parallel} nodes  established equivalent relationships with the overall group, although they do not communicate among them. The strategy of connecting to nodes 1 and 2 or to nodes 33 and 34 is repeated by the children of the parallel nodes. In particular, looking at the subgraph generated by the prime root node (right in Figure \ref{figzachary}), the connecting role of node 1 is more clearly appreciated than in the original graph. It is interesting that vertices 1 and 34 correspond to the figures of instructor and administrator, respectively, which  were the leaders around which the karate club eventually split \cite{zac}.

The MD tree of the \emph{Escherichia coli} transcription factor-gene interaction network has 303 internal nodes and 6 levels. This example is interesting because of the size of the original graph with 1827 nodes and 4.1K edges. Its first level is a \emph{parallel} node with 23 children (connected components). Of these children two are   \emph{prime}, with respectively 4 and  554 children, and the rest are \emph{series} or leaves. The prime nodes have \emph{parallel} or  \emph{series} children, generally with fewer children themselves.  Table \ref{tablanat} shows the overall statistics indicating   almost 8 times as much \emph{parallel} nodes as \emph{series} nodes. In average \emph{series} nodes have 2.05 children, whereas \emph{parallel} have 5.7. \emph{Series} nodes appear at level 2 and 4 and then have \emph{ parallel} children.   \emph{Prime} nodes do not have \emph{ prime } children. In the interpretation of section \ref{MD} (\cite{gagneur}), \emph{series} nodes represent functional units and \emph{parallel} nodes alternative structures. Recall all children of a \emph{parallel} node have exactly the same connections out of the node, but do not communicate among them. Figures are not included because the number of nodes does not allow for an easy interpretation of the resulting MD tree or the original graph.

	\begin{table}
 	\begin{center}
 		{\footnotesize \begin{tabular}{|c|c|c|c|c|c|c|c|}
		\hline \hline
		Network& nodes & edges &	Primes	& Series & Parallel  & \#levels & \makecell{largest \\  prime}  \\ 
			\hline 
		Zachary& 34& 78 & 0.33& 0&0.66	 & 3 & 29 \\ 
		\hline 
		\makecell{Visual cortex \\ mouse 1}& 29& 44 & 	0.25& 0& 0.75	 & 3 &  25\\ 
		\hline
		\makecell{Visual cortex\\ mouse 2}& 193 &214 & 	0.05&  0.00& 0.95	 & 3 &  35\\ 
		\hline\makecell{	Rhesus brain\\ Macaque 1}& 242&4089 & 	0.2& 0& 0.8	 &3 &  234\\ 
		\hline
		\emph{Ec. col.} TF & 1827 & 4348 &0.007& 0.13& 0.86 	 & 6 &  554\\ 
		\hline 
\end{tabular}}
 	\end{center}
\caption{MD statistics for several social and natural networks}
\label{tablanat}
\end{table}

\begin{figure}
	\begin{tabular}{ccc}
		\includegraphics[width=4cm,height=5cm]{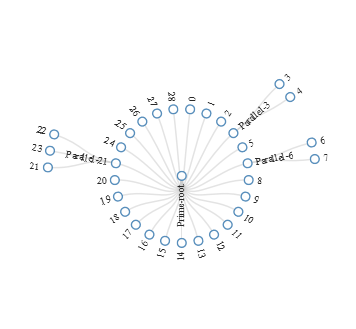}
		&
		\includegraphics[width=4cm,height=5cm]{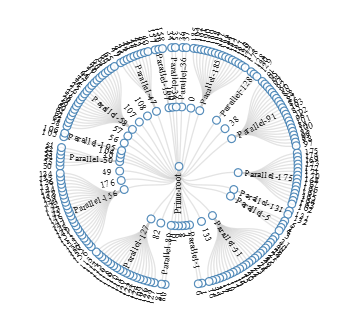}
		&
		\includegraphics[width=5cm,height=5cm]{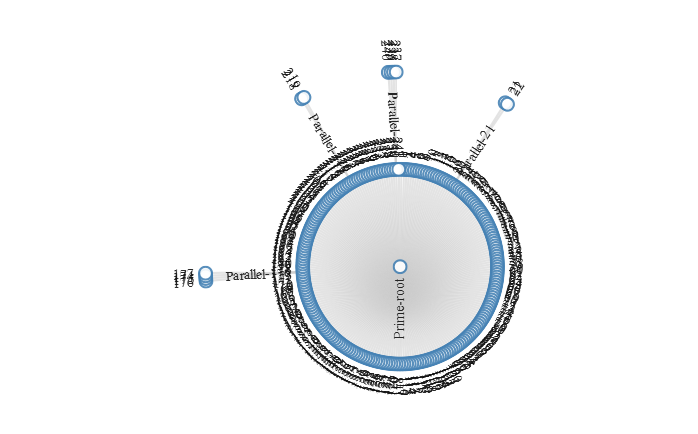}
		
	\end{tabular}
	\caption{ All MD trees have the same structure: the root node is \emph{prime }and second level nodes are \emph{parallel}. Left: MD Visual cortex mouse 1. Center: MD Visual cortex mouse 2 Right: MD brain rhesus macaque (node names are hard to read for the latter, but overall structure is seen to be alike).}
	\label{figbrains}
\end{figure}

In the next examples,  the three brain associated networks for different species (\cite{repository},\cite{ros}) show the same overall behavior \cite{conf1}: a complex first level \emph{prime} along with \emph{parallel} nodes and a 3 level network in each case (see Figure \ref{figbrains} for the resemblance of the overall structure).Statistics for these networks are given in Table \ref{tablanat}. \emph{Series} nodes are absent and  the only prime node to occur in the MD tree is the root. 

\begin{figure}
	  \begin{tabular}{cc}
		
		\includegraphics[width=5cm,height=5cm]{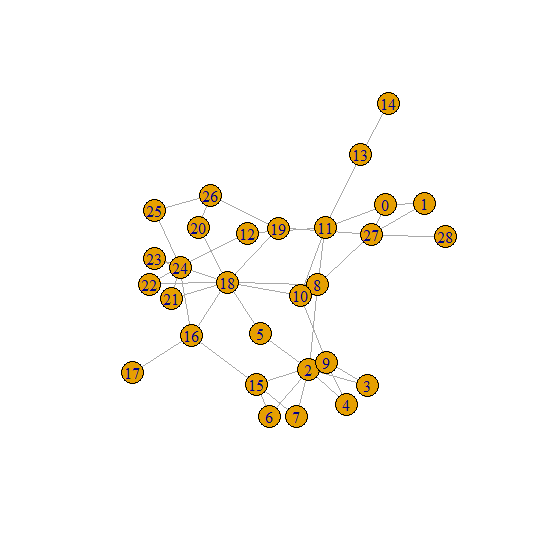}
		&
		\includegraphics[width=5cm,height=5cm]{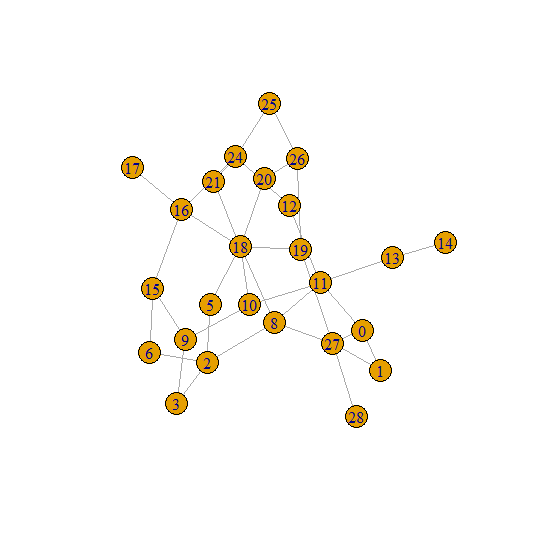}
		
	\end{tabular}
	\caption{  Left:  Visual cortex mouse 1 original graph.  Right: Graph of the prime root node in the MD tree of  Visual cortex mouse 1.}
	\label{figbrains1}
\end{figure}

In the case of the Visual cortex mouse 1 \cite{repository}, because of the small size of the original graph with 29 nodes and 44 edges it is possible to look at the MD tree in more detail as was done with Zachary social network example.  Figure~\ref{figbrains1} compares the original graph to the subgraph generated by the  prime root node in the MD tree. In the original graph, nodes 3 and 4 connect to nodes 2 and 9, nodes 6 and 7 connect to nodes 2 and 15. Finally, nodes 21, 22 and 23 connect to nodes 18 and 24. Connections to nodes 2,9,15, 18 and 24 are repeated by the children of the parallel nodes. The role of node 2 is especially interesting  as connections to this node repeat in two different parallel children of the root.

 This short analysis of the MD tree structure of some selected natural and social graphs, suggests that in general, prime nodes are more probable than complete nodes for connected graphs with a medium to big number of vertices. As the number of vertices decreases series and parallel nodes appear and prime nodes are less probable at the lower levels of the tree. In general, series nodes tend to have a small number of children. Understanding the behavior of the  prime root node when it occurs, seems to be an important issue in describing graph behavior, as well as which children in this prime are parallel or series, and which subgraphs in the original graph are involved. Of course, more studies are required in this direction.

\subsection{Simulated graphs}
We consider simulations of Erd\"os-Renyi (ER) graphs with $n=50$ vertices   and varying link-probabilities and Barab\'asi-Albert (BA) graphs with $n=50$ vertices and ``bag'' method using package Igraph in R (Erd\"os \& Renyi \cite{ER1}, \cite{AB1} and \cite{estrada}, Chapter 11, for a comprehensive review  on random graphs). Mean values and standard deviations for the stated statistics  calculated over $N=50$ simulations are presented for each case in Table \ref{tablesim}. Results remain unchanged by increasing the number of simulations.

\begin{table}[h]
 \begin{center}
 {\small 	\begin{tabular}{|c|c|c|c|c|c|}
 	\hline \hline
 	Simul&	Primes	& Series & Parallel  & Av. \#levels & largest prime \\ 
 	\hline 
 	ER $p=0.01$	&0.02(0.01) & 0.15(0.05) & 0.83(0.05)  & 4 (0.4) &  5.16(1.45)\\ 
 	\hline
 ER	$p=0.05$&	0.18(0.09)&  0.06(0.08) & 0.76(0.1)  & 3.96(2.96) &  42.54(6.17)\\ 
 	\hline ER	$p=0.5$ &	1 (0)& 0& 0	 &2 (0) &  50(0)\\ 
 	\hline
 	BA&	0.13(0.034)& 0 &0.87(0.034) & 3(0) &  37(2.97)\\ 
 	\hline  
\end{tabular}}
 \end{center}
\caption{Simulations for ER and BA methods. Statistics include density of the different types of nodes in the MD tree, average number of levels and average size of longest prime.}
\label{tablesim}
\end{table}

\begin{figure}
	\begin{tabular}{ccc}
		\includegraphics[width=4cm,height=4cm]{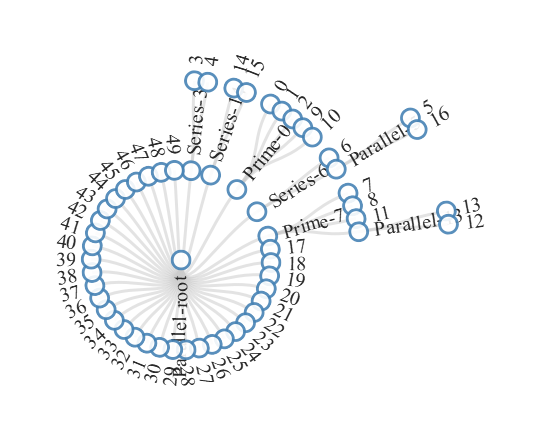}
		&
		\includegraphics[width=4cm,height=4cm]{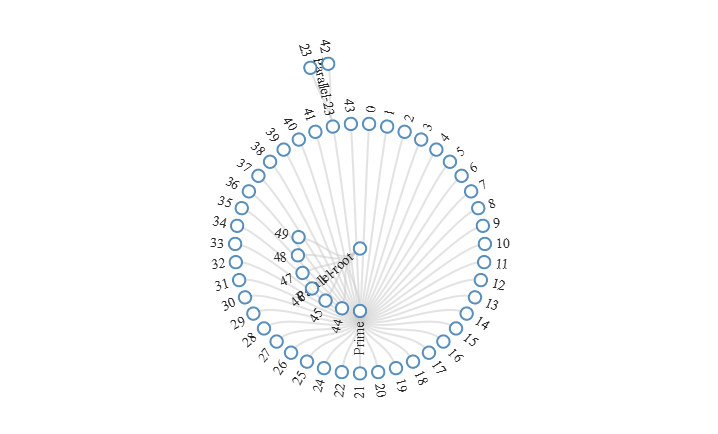}
		&
		\includegraphics[width=4cm,height=4cm]{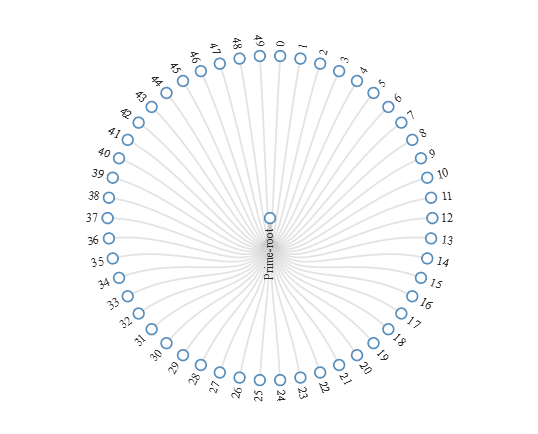}
	\end{tabular}
	\caption{Examples of MD for ER simulated graphs. Left:  $p=0.01$. The first node is \emph{parallel}, followed by only one small \emph{prime} node and the rest are \emph{series} or \emph{parallel}. Center: $p=0.05$. The first node is \emph{parallel}, followed by  one large \emph{prime}  and one small \emph{series}. The \emph{prime} contains an additional small \emph{parallel} node.    Right: $p=0.5$. The first node is \emph{prime} and it contains no further levels.}
	\label{fig-simulER}
\end{figure}

Analysis of the MD tree for simulations with ER and BA random graphs show that the random graph generation procedure has an important effect on the tree structure.
In Figure \ref{fig-simulER} the MD trees for ER graphs with varying link probabilities are presented.  ER simulations with small values of p are not connected, so the root node is parallel. But also, primes tend to be small (with 4 to 6 children) and complete nodes (with 2 children) appear often. As p increases, although typically the root is still parallel, it has a smaller number of children and a big prime node occurs at the second level (this of course is related to the giant component, see for example \cite{estrada}), with much fewer total levels. For big p, the MD tree structure is just a one level tree with a prime node. Thus, smaller values of p yield a complex, larger level MD tree but with small primes. Larger values of p yield simpler tree structures but with large, very connected primes.

BA simulated graphs have an MD tree that is always a prime root node with an average of 37 children which are either  parallel nodes or leaves.  The number of levels was  3 across all simulations. 

\begin{figure}
	\begin{center}
		\begin{tabular}{cc}
		\includegraphics[width=6cm,height=5cm]{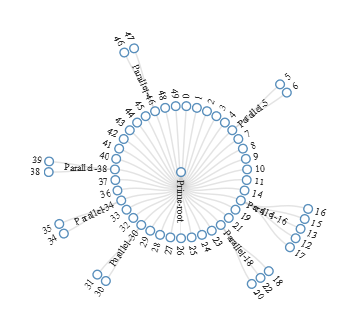}
		&
		\includegraphics[width=6cm,height=5cm]{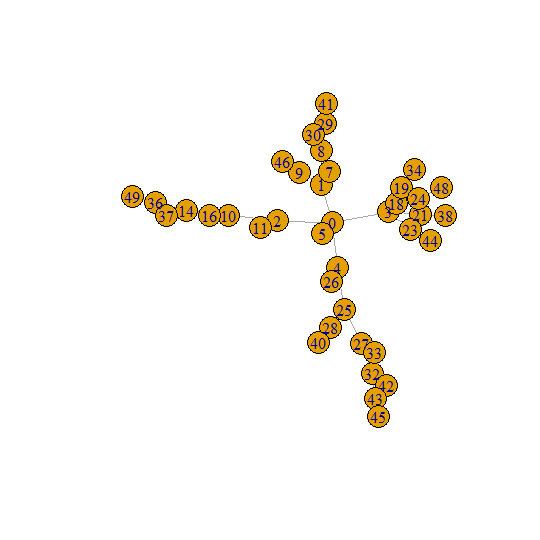}	
	\end{tabular}
	\end{center}
	\caption{Example of MD for BA simulated graph. Left:  MD tree. The first node is \emph{prime}, followed by only  \emph{parallel}  nodes.    Right:  Prime root node. The structure is tree-like.}
	\label{fig-simulBA}
\end{figure}

An example of a simulated BA graph is presented in Figure \ref{fig-simulBA}. The structure coincides with the patterns produced in natural graphs, that is, a prime as the root node whose children are parallel nodes or leaves. However,  the tree structure of the subgraph generated by the prime root node does not resemble neither the subgraph generated by the prime root node in the Zachary MD tree nor the subgraph generated by the prime root node in the visual cortex mouse 1 MD tree. 

Results in this section provide insights for the simulation of graphs based on their MD tree: combining desired probabilities for the root types and the subsequent  appereance  of lower level  nodes. It seems reasonable to assign high probability to prime roots in connected graphs  with a large number of children, followed by parallel or series nodes with fewer children. This shall be discussed in detail in the next section.

\section{A random graph model based on the MD decomposition of a graph}\label{secmdrg}

The proposed random graph model is generative and iteratively  constructs the MD tree of the graph $G$. Each node in the tree is assigned a type, a set of vertices and a number of children.  Starting with  user defined initial and transition probabilities for type and a fixed number of vertices $n=|V|$ the iterative process initializes by creating a root-node $M_0$. The type $T_0\in \{``series",``parallel", ``prime"\}$ is chosen with an initial probability distribution $\pi_0$. A  random variable $K\le n$ is then generated with a user given distribution $F$  to determine the number of initial children of node $M_0$.  Finally, the $n$ vertices of $V$ are arranged among the $K$ children-nodes according to a certain pre-specified rule which we will discuss in detail below. A node which is assigned only one vertex defines a \emph{leaf}. Iteratively, the vertices in each non-leaf child $M_j$ with $n_j>1$ vertices will define a module of the graph $G$ and $M_j$  becomes in turn the root-node of a new tree with its assigned $n_j$ vertices. It is thus assigned a type and number of children. The process stops when all $n$ vertices have been assigned as \emph{leaves}.

Distribution $F$ will be set to depend on the type.  In the case when type is ``prime" then additionally the actual prime with the pre-specifed number of children $K$ is selected at random with a uniform distribution over the set of primes with $K$ vertices.

For a given number of children $K$ and $m$ vertices, the vertices are distributed using a preferential attachment policy (Multivariate Polya Urn Scheme) which can be linear or super linear: starting at time $l=0$ with one vertex for each child, the remaining $m-K$ are assigned to a child  $k\le K$ with probability  $f(m_{l,k})\propto m_{l,k}^\gamma$, $\gamma\ge 1$, where $m_{l,k}$ is the number of vertices  in  child $k$ at time $l$.   For $\gamma=1$ this is the multivariate Polya urn scheme \cite{johnsonykotz}. If $\gamma>1$ it is called a super linear scheme (see for example results in \cite{collevechio}). Conditional on the distribution of vertices,  siblings are assumed independent.

Starting with the first node  $M_0$, iteratively each node $M_j$ depends on two  parameters: its parent type $T_{j-1}$  and its   number of  assigned vertices $n_j$. It is in turn assigned its own type and its  number of children $K_j$ into which the vertices are then distributed. For $j=0$, the parent type is null and the number of vertices is $n$. Given the parent's type, the child's type is defined by the transition matrix

\begin{equation}M_p=\left[\begin{array}{ccc} 0 & p_{ser,par} & p_{ser,pr}\\
p_{par,ser} & 0 & p_{par,pr}\\p_{pr,ser} & p_{pr,par} & p_{pr,pr} \end{array}\right],\end{equation}

\noindent where $p_{i,j}$ is the probability of belonging to type $j$ given the parent is of type $i$. Recall it is not possible to have a  ``series" or ``parallel" child from the same type parent.

The whole process is described in Algorithm \ref{mdrg}

\begin{algorithm}\label{mdrg}
	\KwData{Number of vertices $n$, distribution  $F$ and its parameter list,  parameter $\gamma$, initial type distribution $\pi_0$ and transition matrix $M_p$}
	\KwResult{Graph $G:=(V_j,T_j,K_j)_{j=1}^J$, the adjacency matrix $A$ and list $L$ indicating for each vertex the node where it is a leaf}
	Initialization: Set $j=0$. Set $V_j=\{1,\ldots,n\}$. Generate $T_j$ with $\pi_0$ and $K_j\sim F$.   Set $S=\emptyset$ ($S$ is the set of leaves)\;
	\While{$|S|<n$}{Generate a partition of $V_j$, $v_1,\ldots,v_{K_j} $ using a preferential attachment scheme with parameter $\gamma$. Set $n_{j,k}=|v_k|$.\;
		
		\For{$k\le K_j$}{
			$jaux=j$\;
			\If{$n_{j,k}>1$}{
			$j=j+1$\;
			$T_j\sim p_{T_{j-1},\cdot}$
			$K_j\sim F$\;
			$V_j=v_k$
			\If{($T_j=$``prime")}{Generate $P_j$ prime of size $K_j$ with uniform distribution\;} 
			Update $A$  for component $j$\;		
		}
	\If{$n_{jaux,k}=1$}{
		$S=S\cup v_k$\; 
	$L(v_k)=jaux$}
	}
}
	\caption{MD Random graph generator}
\end{algorithm}
General considerations:
\begin{enumerate}
	\item The distribution $F$ determines the form of the generated graph as will be shown in simulations. The following are possible examples:
	\begin{enumerate}
		\item Truncated Poisson with parameter $\lambda$
		\item Uniform $\{2,\ldots, K'\}$, for some fixed $K'$
		\item Truncated power law with exponent $\alpha$: \begin{equation}\label{truncpareto}
		P(K=k)=k^{-\alpha}/\sum_{k=k'}^{K'} k^{-\alpha},\, k'\le k\le K'
		\end{equation}
		\item Truncated power law as in (\ref{truncpareto}) for prime  types and Uniform $\{2,\ldots, K'\}$ for series or parallel types. 
	\end{enumerate}
	\item The first node plays an important role for many properties of the generated graph. In particular, sometimes it might be reasonable to pre-specify whether the first type is parallel or not as that will define whether the generated graph is connected or not. This can be defined by user defined distribution $\pi_0$.
	\item As seen in the examples of Section \ref{exps}, typically parallel nodes tend to appear more often and have a medium number of children, series appear less often than parallel nodes, with 2 or 3 children, whereas prime nodes tend to appear few times and have a larger number of children. Also, probability of primes tends to decrease with the level of the MD tree. We model this by prohibiting primes when the number of vertices is smaller than a given threshold.  It may be of interest to consider parametric models with decreasing probability of primes as a function of the level.

\end{enumerate}

\subsection{Properties}
The hierarchical nature of the proposed model allows for a very interesting level-wise approximation to the global graph properties.  Although precise mathematical results for the random graphs generated with Algorithm \ref{mdrg} are out of the scope of this article, there are  several  global properties which occur naturally from the definition and which we describe in what follows.  We start by introducing some necessary notation and previous results:
\begin{enumerate}
	\item The graph is defined by the nodes $M_j$ arranged in the MD tree. Each node has an assigned type $T_j$ and a collection of $n_j$ vertices arranged in a random number $K_j$ of children. Vertices in child $k$ of node $M_j$ are called $V_{j,k}\subset V$. Define $V_j=\cup_k V_{j,k}$ Set $n_{j,k}=|v_{j,k}|$. By construction, $T_j$ and $n_j$ only depend on $T_k$ and $n_k$ of node $M_k$ if $V_j\subset V_k$ and there does not exist $V_m$ with $V_j\subset V_m\subset V_k$.
	\item Each $v\in V$ is assigned the path $\mbox{path}(v):=\{M_{j_i}\}_{i\ge 0}$ of nodes containing $v$ from the the root $M_0$ to the leaf $v$, such that $M_{j_0}=M_0$ is the first level node. Set $L(v):=|\mbox{path}(v)|$ to be the length of $\mbox{path}(v)$. The sequence ends at the node $M_{L(v)}$ for which the child that contains $v$ contains no other vertex ($v$ is \emph{leaf}). We also define $K_j(v)$ as the child containing vertex $v$ in $M_j$ and $N_j(v)$ as the indexes of the neighbors of $K_j(v)$ in $M_j$. Clearly, $N_j(v)$ depends on $T_j.$ Set $|N_j(v)|$ to be the number of neighbors of $v$ at level $j$. 
	\item When $\gamma=1$, allocation in children is a Multivariate Polya-Eggenberger distribution given the number of children $k$ and number of vertices $n$,  with $s=1$ and $c_j=1,j=1,\ldots, k$ in the notation of \cite{johnsonykotz}, pg. 194. Thus,  the marginal distribution of the number of vertices in any given child follows a Polya-Eggenberger distribution (unidimensional). Also then,  the expected number of vertices in any child is $E(n_{j,k})=n_j/K_j$, the variance is $Var(n_{j,k})=n_j(K_j+n_j)(K_j-1)/(K_j^2(K_j+1))$ and the covariance is $Cov(n_{j,k}n_{j,l})=n_j(n_j-1)/(K_j(K_j+1))$ (\cite{johnsonykotz} pgs 194-195). As a consequence, adding levels can be thought of as simply adding  children. Given a sequence $(K_1,K_2,\ldots, K_j)$ of number of children along the path of any given vertex, the expected number of vertices in any child $k$ at level $j$ will be $n/(K_1+\cdots K_j)$. Moreover, given the path $\mbox{path}(v)$, at each level $j$, vertices may be grouped in two components: $K_j(v)$ and $K_j^c(v)$. Let $|A|$ stand for the number of vertices in any child $A$.  Since the last descendant of  $\mbox{path}(v)$ has one vertex by construction, this means that $\sum_{j=1}^{L(v)} |K_j^c(v)|=n-1$ and $|K_j(v)|>1$ for any $j<l$.  Then the event $\{L(v)=l\}$ is equal to the event $\{l=\min_t\sum_{j=1}^{t} |K_j^c(v)|=n-1\}$ which in turn corresponds to all ways of distributing $n-1$ balls in $l$ urns with no empty urns. Whence, $p(L(v)=l)=\left(\begin{array}{c} n-2\\ l-1 \end{array}\right).$
\end{enumerate}
With the above notation and results for $\gamma=1$ we are able to give some preliminary results regarding the behavior of some graph characteristics: namely the structure of its adjacency matrix, degree distribution, diameter and clustering coefficient. 
\begin{enumerate}
	\item Adjacency matrix:  by construction, the adjacency matrix $A:=[a_{i,j}]$ of the generated graph has an iterative block structure. Recalling a module $M$ is defined by the property that $a_{i,j}$ is constant for all $v_i\in M$ and $v_j\not\in M$, each node $M_j$ generates an outer graph over its $K_j$ children. Let $A_1$ be the adjacency matrix of module $M_1$ over its $K_1$ children, so that,
		\begin{equation}A_1=\left[\begin{array}{ccc}0&\cdots&b_{1,K_1} \\
		&\ddots&\\
		b_{K_1,1}&\cdots&0
	\end{array}\right]\end{equation}
	where each $b_{i,j}\in \{0,1\}$ indicates whether child $V_{1,i}$ is connected or not to child $V_{1,j}$.
	 Each $V_{1,k}$ in turn generates an adjacency matrix $A_{1,k}$ of size $n_{1,k}$. Thus using this first level decomposition, we can write
	\begin{equation}
	\label{block}
	A=\left[\begin{array}{ccc}A_{1,1}&\cdots&B_{1,K_1} \\
		&\ddots&\\
		B_{K_1,1}&\cdots&A_{K_1,K_1}
		\end{array}\right]\end{equation}
		where each $B_{i,j}$ is a constant block matrix of size $n_{1,i}\times n_{1,j}$ with entries $b_{i,j}$. Iterating the above procedure for each  $A_{1,k}$ yields the stated block representation. For  applications such as degree calculation or triangle counting this iterative block structure gives a level-wise approximation strategy which provides interesting insight about the overall structure.
	\item Degree of a vertex and degree distribution:  by construction the degree of each vertex $v$ may be defined as 
\begin{equation} \label{deg}dg(v)=\sum_{M_{j}\in \mbox{path}(v)}\sum_{k\in N_j(v)  } n_{j,k}.\end{equation}
Alternatively, equation (\ref{deg}) can be obtained form the block decomposition described in (\ref{block}) and calculating the diagonal of $A^2$, using $deg(v)=A^2(v)$.
	Whence, degree distribution depends on the collection $\{n_{j,k}\}$. As described, the model assumes a multivariate linear ($\gamma=1$) and non-linear ($\gamma>1$) Polya distribution for vertices given the number $K$ of children. General results are complex. We limit our theoretical exposition to the linear case.
	We have the following Lemma
	\begin{lemma}\label{lemdeg} Assume a random graph is generated using Algorithm \ref{mdrg} with $\gamma=1$. Then,
		\begin{eqnarray}\label{expec}
	E(dg(v))&=&\sum_l P(L(v)=l)\\ \nonumber
	&\times& \sum_{j=1}^l \sum_{K_1,\ldots, K_j} E(n_{j,k}|L(v)=l,K_1,\ldots,K_j)E(|N_j(v)|\,\,|k_j)\\ \nonumber
	&\times& P(K_1,\ldots,K_j)
		\end{eqnarray}
		In particular, using a first level approximation, \\
			\begin{eqnarray}\label{expec1}
		E(dg(v))&\ge &
		 \sum_{k} \frac{n}{k} E(|N_1(v)|\,\,|k) P(k)
		\end{eqnarray}
		Also, the following inequality in probability holds 
		\begin{equation} \label{prob}P(dg(v)>m)>\sum_{k=m}^{n-k}P( K_1=k)P(|N_1(v)|=m|k)\end{equation}
		And, in particular, if the number of children follows a truncated power distribution, then
		\begin{equation} \label{prob1}P(dg(v)>m)> C(\alpha)\sum_{k=m}^{K'}k^{-\alpha} P(|N_1(v)|=m|k)\propto m^{-\alpha+1} \end{equation}
	\end{lemma}
\begin{remark}
	Without the constraint $L(v)=l$, $E(n_{j,k}|K_1,\ldots,K_j)=\frac{n}{K_1+\cdots+K_j}$, as follows by just adding new ``urns''. 
\end{remark}
\begin{remark}Decomposing into the possible types, 
	\begin{eqnarray*}
	&&E(|N_j(v)|\,\,|k_j)\\&=&K_j(K_j-1)P(series)+E(|N_j(v)|\,\,|k_j,T_j=prime)P(prime).
	\end{eqnarray*}
Where $E(|N(v)|\,\,|k,T=prime)$ is actually $E(deg(v))$ for a prime graph ($l=1$ and $T_1=prime$). Sometimes it is interesting to allow for a big probability first level parallel module, which would lower the r.h.s of (\ref{expec1}) and (\ref{prob1}). For practical applications. w.l.o.g. it can be assumed that we are considering models such that $P(T_1=\mbox{parallel})=0$ since this amounts to modeling connected components.
\end{remark}
\begin{remark}
	Inequality (\ref{prob1}) holds when the number of children follows a power distribution, thus favoring existence of nodes with a large amount of children. In practice, large \emph{series} nodes are highly improbable, so that the result suggests existence of big \emph{prime} nodes.
\end{remark}
\noindent Proof:
We begin with the proof of (\ref{expec}). From (\ref{deg}),
\begin{eqnarray}E(dg(v))&=&E(\sum_{M_{j}\in \mbox{path}(v)}\sum_{k\in N_j(v) }n_{j,k})\\ \nonumber
& =& 	\sum_l P(L(v)=l)\sum_{j=1}^l E(\sum_{k\in N_j(v) }n_{j,k}|L(v)=l)\\ \nonumber
&= & 	\sum_l P(L(v)=l)\sum_{j=1}^l\sum_{K_1,\ldots,K_j}P(K_1,\ldots,K_j)\\ \nonumber
&\times&	E(\sum_{k\in N_j(v) }n_{j,k}|L(v)=l,K_1,\ldots,K_j)
\end{eqnarray}
Now, for each $j$, and $k$, since the addition of extra levels with $K_j$ children each simply increases the number of urns in which to distribute the n vertices, we have
$E(n_{j,k}|L(v)=l,K_1,\ldots,K_j)$ does not depend on $k$ or on the particular sequence
$n_j$.
On the other hand, for a given level $j$, since the number of neighbors only depends on the type $T_j$ and number of children $K_j$ and the expectation of the number of vertices in each child is independent of the number of neighbors, we have
\begin{eqnarray}
 &&E(\sum_{k\in N_j(v) }n_{j,k}|L(v)=l,K_1,\ldots,K_j)\\ \nonumber&=&\sum_s P(N_j(v)=s|L(v)=l,K_1,\ldots,K_j)\\\nonumber &\times&\sum_{k=1}^s E(n_{j,k}|L(v)=l,K_1,\ldots,K_j,s)\\ \nonumber
&=&\sum_s P(|N_j(v)|=s|K_j)sE(n_{j,k}|L(v)=l,K_1,\ldots,K_j)\\ \nonumber
&=&E(|N_j(v)|\,\,|K_j)E(n_{j,k}|L(v)=l,K_1,\ldots,K_j,s)
\end{eqnarray}
Inequality (\ref{expec1}) then follows by setting $l=1$ and using that $E(n_{1,k}|k)=\frac{n}{k}$
For the proof of (\ref{prob}), we begin again with (\ref{deg}). The event $deg(v)\ge m$ contains the event $$E=\{\mbox{the first level contains $l>m$ children of which m $\in N_1(v)$} \}.$$ Thus
$P(deg(v)\ge m)\ge \sum_{k\ge m} P(K_1=k)P(|N_1(v)|=m|k)$. The bound in (\ref{prob1}) follows by	assuming $K$ follows a truncated power  distribution.
	\hfill\qed
	\item Small diameter: the distance among all vertices belonging to any child of a prime or series node is at most two. On the other hand, the diameter of the subgraph defined by the vertices in any given module is that of the outer. Thus, the existence of a large node of prime or series type assures small average diameter. We have the following result
	\begin{lemma}For any connected graph, let $K$ be the number of children of the first level
		\begin{equation}E(diam)\le 2 P(series)+ P(prime)*E(K)\end{equation}
	\end{lemma}
Proof:
Conditional on the fact the graph is connected, the first node must be either prime or series. In the latter, all vertices are separated by at most 2, since vertices in any one child are all connected to all vertices in another child and all children are connected. In the former, as before, all vertices in one child are separated by at most two, and if the number of children is $k$, the max distance among chidren is $k$. Thus
\begin{eqnarray}
& &E(diam)\le \\ \nonumber & &2P(series)+P(prime)*\sum_k k P(\mbox{number of children}=k) 
\end{eqnarray}
which yields the stated result.\hfill\qed
	\item Clustering coefficient (local): given a vertex $v$ its Watts-Strogatz clustering coefficient is defined as $C(v)=\frac{2t(v)}{dg(v)(dg(v)-1)}$ (for example, see \cite{estrada}, pg. 101), where $t(v)$ is the number of triangles vertex $v$ belongs to and $dg(v)$ is its degree. Given a level $j$, let $t_j(v)\subset N_j(v)$ denote the neighbors of $v$ which are also neighbors, and $|t_j(v)|$ the number of vertices in this set. By construction,  the number of triangles for vertex $v$ is
	\begin{equation} \label{tri}t(v)=\sum_{M_{j}\in \mbox{path}(v)}\sum_{k,l\in t_j(v) } n_{j,k}n_{j,l}.\end{equation}
	Alternatively, equation (\ref{tri}) can be obtained from the block decomposition described in (\ref{block}) and calculating the diagonal of $A^3$, using that $t(v)=A^3(v)$. Although a formal characterization of $E(C(v))$ is out of the scope of this article, some insights are possible for the Polya Urn scheme with $\gamma=1$. The following result is useful
		\begin{lemma}\label{lematriang} Assume a random graph is generated using Algorithm \ref{mdrg} with $\gamma=1$. Then,
			\begin{eqnarray}\label{triang1}
			E(t(v))&=&\sum_l P(L(v)=s)\\ \nonumber
			&\times& \sum_{j=1}^s \sum_{K_1,\ldots, K_j} E(n_{j,k}n_{j,l}|j\ne l,L(v)=s,K_1,\ldots,K_j)\\ \nonumber &\times&   E(|t_j(v)|\,\,|K_j)
			P(K_1,\ldots,K_j)
			\end{eqnarray}
	\end{lemma}
Proof: it follows exactly as the proof of Lemma \ref{lemdeg}.
\hfill\qed

In the definition of $C(v)$, the product in its denominator can be calculated using equation (\ref{deg}).   We restrict the analysis of $C(v)$ to a first level approximation, conditioning on the number of children $K_1$ and $n$ vertices. Let $n_k$ be the number of vertices in child $k$.  The covariance to variance ratio of the number of vertices in the children, given $K$, is then
\begin{equation}r(n,K):=Cov(n_k n_l|K)/Var(n_k|K)=\frac{(n-1)K}{ (K-1)(K+n)}.\end{equation} 
The expectation of the ratio  
\begin{equation}R_1=\frac{2\sum_{k,l \in t_1(v)} n_{1,k}n_{1,l}}{\sum_{k \in N_1(v)} n_{1,k}(n_{1,k}-1)}\end{equation} 
can then be approximated by the ratio of the  conditional expectations given the number of children of the first level  as
\begin{equation}E(R_1)\approx \sum_k P(K_1=k) \frac{2E(t_1|k)}{E(N_1|k)}r(n,k)=E_K\frac{2E(t_1|K)}{E(N_1|K)}r(n,K),\end{equation}
where $E_K$ stands for expectation with respect to the number of children $K$.
For $T_1=series$, $\frac{2E(t_1|k)}{E(N_1|k)}=1$. For $T_1=prime$ this ratio can be also very high, but no theoretical results are available. 
\end{enumerate}

\section{Simulations}\label{simul} 
Our simulations are based on Algorithm \ref{mdrg}. For simplicity and because of space limitations we only discuss scheme (d) in 1, pg. 7, with the following characteristics
\begin{enumerate}
	\item Initial only prime non zero probability ($\pi_0=(0,0,1)$): if $n>4$ the initial module will be  \emph{series} with very low probability and the \emph{parallel} case is just looking at connected components.
	\item Transition probabilities favor  passing from \emph{prime} to \emph{parallel} nodes, prohibit \emph{parallel-parallel} or \emph{series-series} transitions and finally favor \emph{series-parallel} and \emph{parallel-series} transitions. We also consider  the transition matrix $M_p$ for two separate cases: when the number of vertices in a given node is smaller than $K_v=6$, \emph{prime} nodes are not allowed. Thus the transition matrix in this case only allows changing from the \emph{parallel} to the \emph{series} case or viceversa. If the number of children is larger than $K_v=6$ then the considered matrix is 
	$$M_p=\left[\begin{array}{ccc} 0 & 0.7 & 0.3\\
	0.2 & 0 & 0.8\\0 & 0.95 & 0.05 \end{array}\right],$$
	\item \emph{Series} nodes are only allowed $K=2$ children. For \emph{parallel} nodes the number of children is assumed to follow a uniform distribution on $\{2,\ldots,K_v\}$, with $K_v=6$. For \emph{prime} nodes, with $n$ vertices, $n>6$, the number of children is assumed to follow a truncated Pareto distribution:
	$P(K=k)=k^{-\alpha}/\sum_{l=6}^n l^{-\alpha}$.
	\item \emph{Prime} nodes are created using the ER model with $p=0.5$, that is, the uniform distribution over the set of all graphs with $n$ vertices, and checking that the graph is prime using the MD of the graph (only one \emph{prime} node). This procedure is  an acceptance-rejection method. More precisely, let $\mathcal{G}_n$ stand for the set of all graphs with $n$ vertices and $\mathcal{P}_n$ stand for the set of all prime graphs with $n$ vertices. For a given set $A$, let $|A|$ equal the number of elements of $A$ and $1_{x\in A}$ be the logical function equal to 1 if $x\in A$ and 0 otherwise. With this notation, $f(g)=1_{g\in \mathcal{P}_n}/|\mathcal{P}_n|$ is the uniform distribution over the set of prime graphs with $n$ vertices and $h(g)=1_{g\in \mathcal{G}_n}/|\mathcal{G}_n|$ is the uniform distribution over the set of graphs with $n$ vertices. We have $h(g)/f(g)<|\mathcal{G}_n|/|\mathcal{P}_n|:=c$ and for each proposed $g\in \mathcal{G}_n $, $g$ is selected with probability 1 if $g\in \mathcal{P}_n$ or is not selected if $g\not \in \mathcal{P}_n$. As discussed in Section 2, prime graphs density tends to one, so that $c\to 1$ and the procedure is efficient.
	\item Changing values of parameter $\alpha$  the cases $\alpha=0.08$, $\alpha=0.1$ and $\alpha=1$ are considered. 
\end{enumerate} 

\begin{figure}
	\begin{tabular}{ccc}
		\includegraphics[width=4cm,height=4cm]{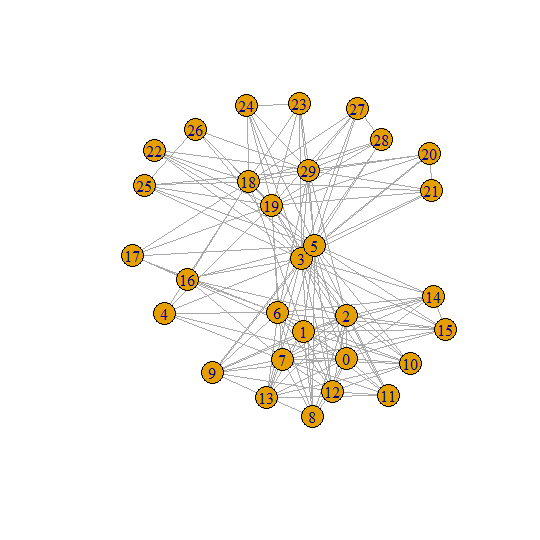}
		&
		\includegraphics[width=4cm,height=4cm]{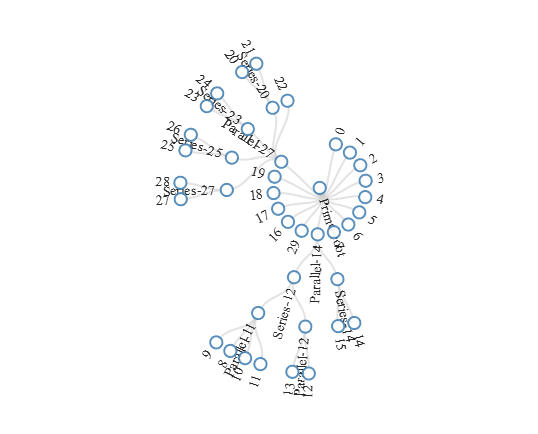}
		&
		\includegraphics[width=4cm,height=4cm]{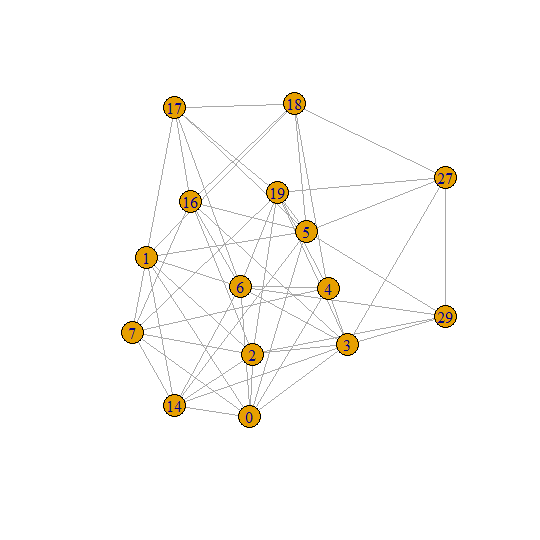}
		\end{tabular}
	\caption{ Left:  Simulated graph with $n=30$ vertices and $\alpha=0.1$.   Center: MD. The first node is \emph{prime} and the rest are either \emph{parallel} or \emph{series}. Right: prime root node (reproduces natural graph behavior).}
	\label{grafosimul}
\end{figure}

A simulated graph showing both the MD, the original graph and principal prime for $n=30$ and $\alpha=0.1$ is shown in Figure \ref{grafosimul}. The root prime resembles the root primes obtained for natural networks presented in Section \ref{exps}.

Based on $N=50$ simulations, Table \ref{table11} shows mean and sd of density, diameter,  global clustering coefficient (3x number of triangles/number of all triplets open or closed) and average local clustering coefficient (Watts-Strogatz) calculated using package Igraph in R.

	\begin{table}[h]
\begin{center}
{\small 	\begin{tabular}{|c|c|c|c|c|}
	\hline \hline
	$\alpha$	& Density & Diameter &  Clusterization (global) & Clusterization (local) \\ 
	\hline 
	0.08& 0.45(0.09) &2.09(0.3) &0.47(0.06)  &0.52(0.06)  \\ 
	\hline 
	1& 0.40(0.11) &2.28(0.45) &0.46(0.06)  &0.54(0.07)  \\ 
	\hline 
\end{tabular}}
\end{center}
\caption{Statistics for simulated graphs using Algorithm 1}
\label{table11}
\end{table}

\begin{figure}[h]
	\begin{tabular}{ccc}
		\includegraphics[width=4cm,height=5cm]{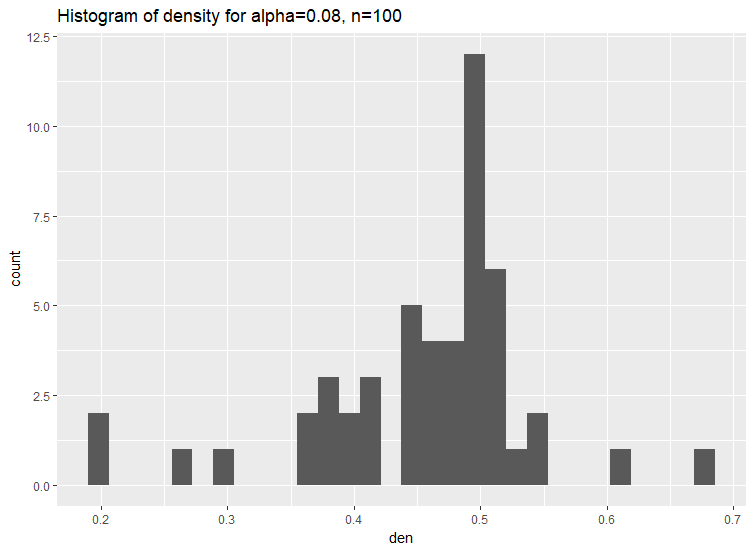}
		&
		\includegraphics[width=4cm,height=5cm]{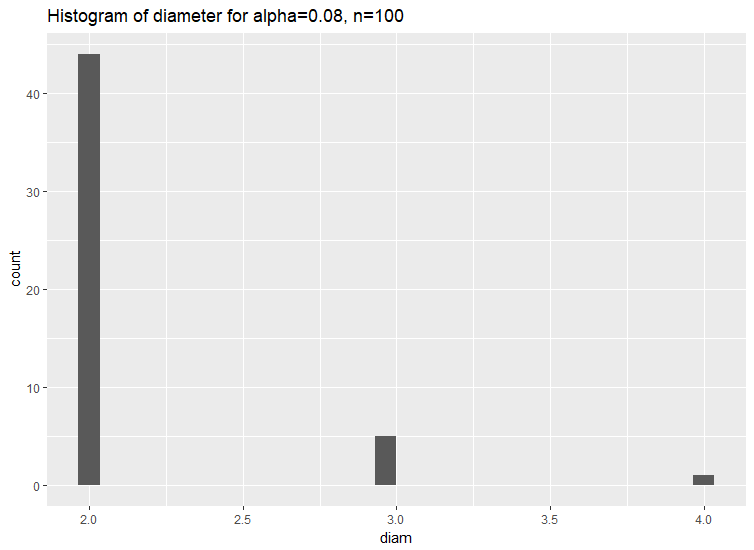}
		&
		\includegraphics[width=4cm,height=5cm]{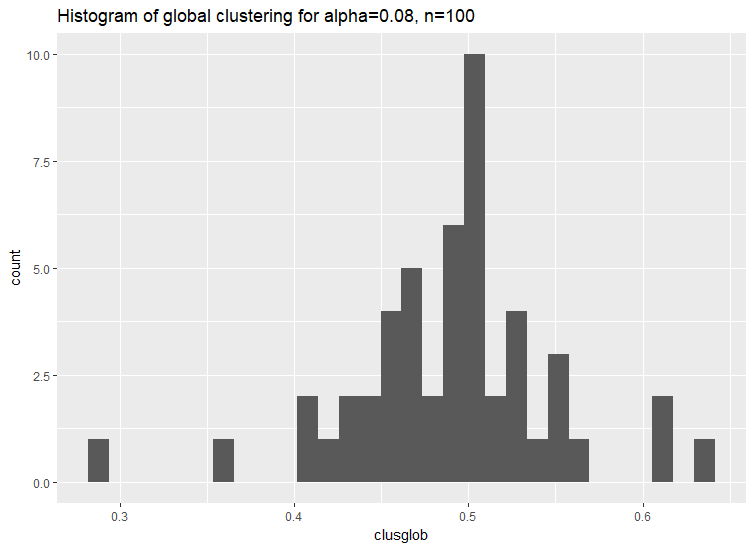}
	\end{tabular}
		\caption{Histograms of density (left), diameter (center) and global clustering coefficient (right) for $N=50$ graphs simulated for $n=100$ vertices and $\alpha=0.08$}
		\label{hists1}
\end{figure}

\begin{figure}[h]
	\begin{tabular}{ccc}
		\includegraphics[width=4cm,height=5cm]{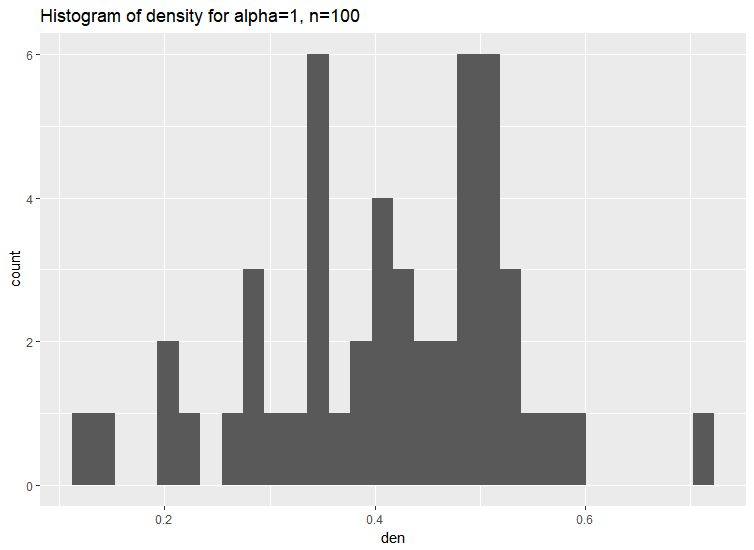}
		&
		\includegraphics[width=4cm,height=5cm]{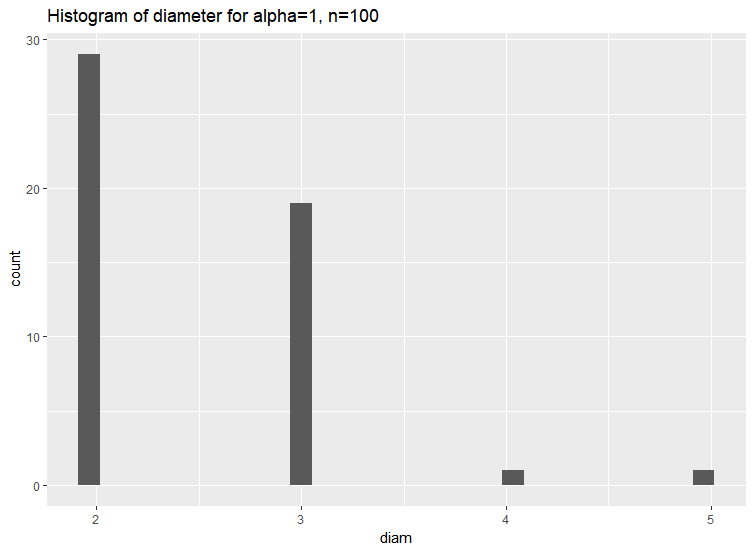}
		&
		\includegraphics[width=4cm,height=5cm]{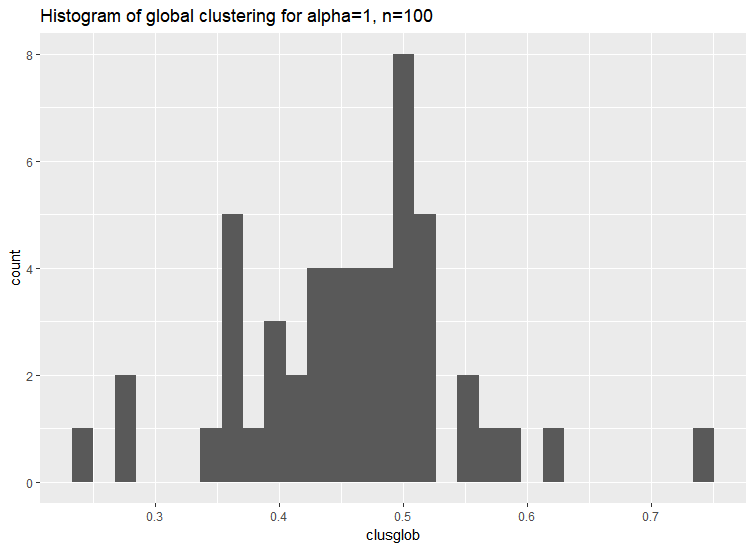}
	\end{tabular}
	\caption{Histograms of density (left), diameter (center) and global clustering coefficient (right) for $N=50$ graphs simulated for $n=100$ vertices and $\alpha=1$}
	\label{hists2}
\end{figure}

\begin{figure}[h]\label{fig-deglolog}
	\begin{tabular}{cc}
		\includegraphics[width=6cm]{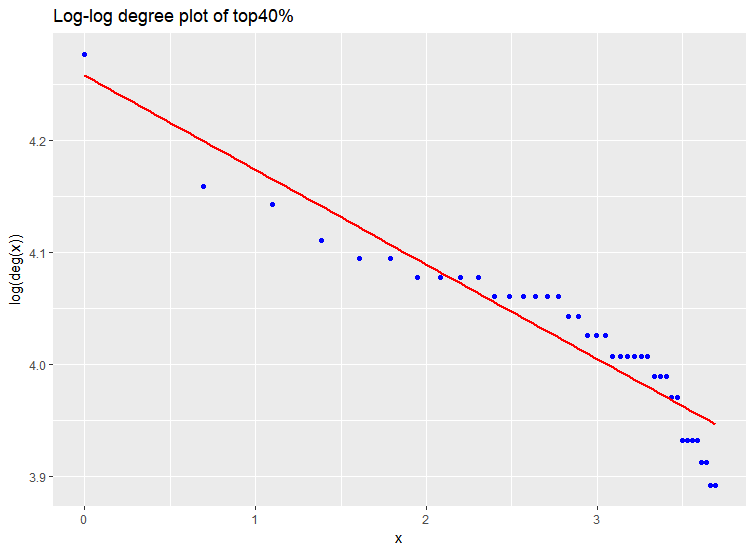}
		&
		\includegraphics[width=6cm]{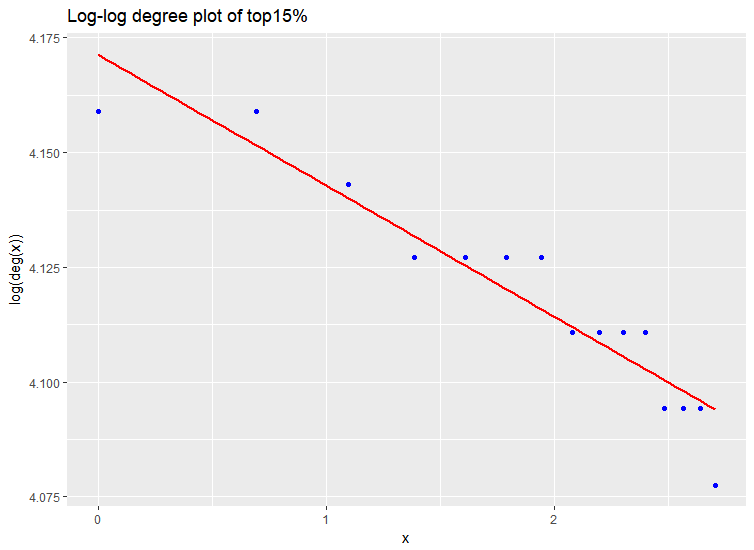}
	\end{tabular}
	\caption{Log-log plots of highest degrees. Left: $n=100,\alpha=1$. Right: $n=100,\alpha=0.08$}
\end{figure}

\begin{figure}[h]\label{fig-deg}
	\begin{tabular}{cc}
		\includegraphics[width=6cm]{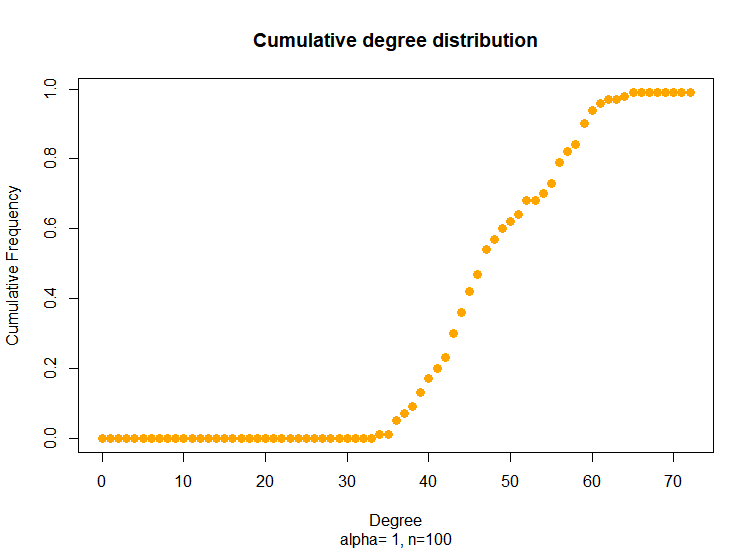}
		&
		\includegraphics[width=6cm]{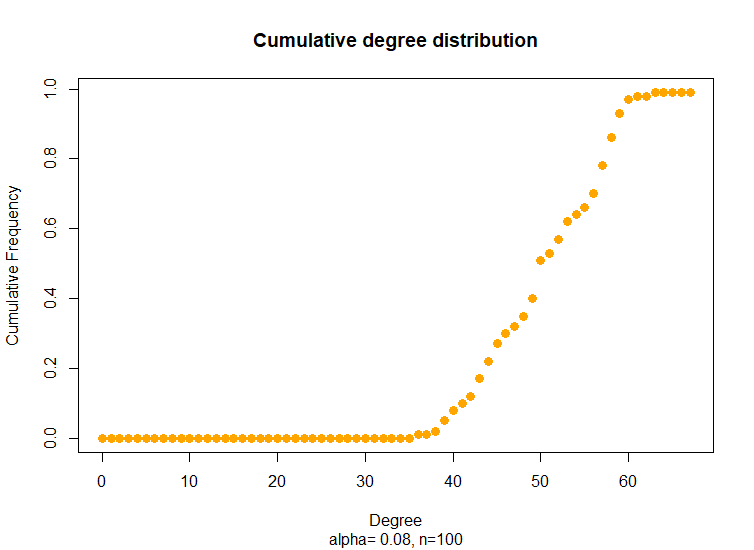}
	\end{tabular}
	\caption{Cumulative distributions of degree distribution. Left: $n=100,\alpha=1$. Right: $n=100,\alpha=0.08$}
\end{figure}

 Figures \ref{hists1} and \ref{hists2}  show distributions for edge density, diameter, average distance, global and average local clustering coefficient for $n=100$ and $\alpha=0.08$ and $\alpha=1$ respectively.
On the other hand, degree distribution follows a power law for largest degrees as can be inferred from Figure \ref{fig-deglolog}. Smaller degrees tend to appear with high frequency in the considered simulation scheme. If $\alpha$ is larger, less small degree vertices appear, but degrees tend to grow by steps as seen in Figure \ref{fig-deg}.
\section{Concluding remarks}\label{conc}
In this article we have used the concept of Modular decomposition (MD) of discrete structures applied to graphs in order to help understand the hierarchical local nature of complex networks. In a sense, MD can be interpreted as a generalization of the idea of connected components, allowing for the possibility of connections  in a such a way that they are the same across components, which are then  termed \emph{modules}.

This decomposition is applied to both natural and simulated networks in order to understand how this hierarchical decomposition behaves in practice and relate this behavior to simulations. Connected graphs tend to have a first level \emph{prime} node with a big number of children, followed by smaller \emph{parallel} or \emph{series} nodes with fewer children. \emph{ Series} nodes seem to appear less, and their size in general is smaller.

Finally, we introduce a novel hierarchical random graph model, based on the fundamental property that any graph has an associated \emph{unique} MD. We give theoretical and simulated results for graphs generated according to the proposed model, showing several interesting properties such as scale free behavior, small  average diameter and large global and average clustering. These results are promising and we hope that further more detailed theoretical and experimental results will  enrich the potential of this proposed method as well as applications to understanding natural and social networks.

\section{Funding}
\noindent \emph{This work was partially supported by project 3991 of Banco de la República, Colombia.}
\section{ Aknowledgments}
\noindent \emph{ The authors wish to thank Diego Villamizar for the development of the MD code in java.}


\clearpage

\begin{thebibliography}{aaaaaa}
	
	
		\bibitem{AB1}A.L. Barabasi \& R. Albert. (1999) Emergence of scaling in random networks, \emph{Science}, 286:509--512.
		
		\bibitem{repository} D. Bock, ET. AL. (2011) Network anatomy and in vivo physiology of visual cortical neurons.  \emph{Nature}, 471, pp. 177--182. 
		
	\bibitem{AB2}B. Bollobas \& O. Riordan. (2003) \emph{Mathematical results on scale-free random graphs}, in Bernholdt, S.and Schuster, H.G. (eds.), Handbook of Graph and Networks: From the Genome to the Internet, Wiley-VCH,pp. 1--32.
	
	\bibitem{collevechio}  A. Collevecchio ET. AL. (2013)  On a preferential attachment and generalized Pólya's urn model. \emph{The Annals of Applied Probability}, Vol. 23, No. 3, pp. 1219--1253.
	\bibitem{regulonDB}Gama-Castro S ET. AL. (2016) "RegulonDB version 9.0: high-level integration of gene regulation, coexpression, motif clustering and beyond.", \emph{Nucleic Acids Res.}, 2016 Jan 4;44(D1):D133-43. 
	
	\bibitem{mdalg} A. Ehrenfeucht ET. AL. (1994) An $O(n^2)$ Divide and Conquer Algorithm for the Prime Tree Decomposition of Two-Structures and Modular Decomposition of graphs. \emph{Journal of Algorithms}, 16:283--294.
	\bibitem{ER1} P. Erd\"os \& A. Renyi. (1959) A. On Random Graphs I.   \emph{Publicationes Mathematicae (Debrecen).} Volume, 6, pp. 290--297.  
	
	\bibitem{estrada}E. Estrada \& P. Knight.  (2015) \emph{A First Course in Network Theory.} Oxford University press.
		\bibitem{gagneur}J. Gagneur ET. AL. ( 2004)  Modular decomposition of protein-protein interaction networks. \emph{Genome Biology}, Vol. 5, Issue 8:R57. 
		\bibitem{gallai} T. Gallai. (1967) Transitive orientabare graphen.\emph{ Acta mathematica Hungarica}, 18, pp.25--66.
		
				\bibitem{habib} M. Habib ET. AL. (2004) A Simple Linear-Time Modular Decomposition Algorithm for Graphs, Using Order Extension. In: Hagerup T., Katajainen J. (eds) Algorithm Theory - SWAT 2004. SWAT 2004. \emph{Lecture Notes in Computer Science}, vol 3111. Springer, Berlin, Heidelberg.
				\bibitem{han}M. S. Handcock,  A. E. Raftery \& J. M. Tantrum. (2007) Model-based clustering for social networks. \emph{J. R. Statist. Soc. A,}
				170, Part 2, pp. 1--22
	\bibitem{egms1} P. Holland \&  S. Leinhardt. (1981) An Exponential Family of Probability Distributions for Directed Graphs. \emph{Journal of the American Statistical Association}, Vol. 76, No. 373, pp. 33--50.
	\bibitem{SBM1} P. Holland et. al. (1983) Stochastic Blockmodels: First Steps. \emph{Social Networks - SOC NETWORKS.} 5, pp. 109--137.
	
	\bibitem{johnsonykotz}   N. Johnson \&  S. Kotz. (1977)  \textit{Urn models and their applications: An approach to modern discrete probability theory},  Wiley.
\bibitem{karwa} Karwa, V., Pelsmajer, M. J., Petrovic, S., Stasi, D., \& Wilburne, D. (2017). Statistical models for cores decomposition of an undirected random graph. Electronic Journal of Statistics, 11(1), 1949-1982.

	\bibitem{largegraphs} A. Mbaya \&  O. Hammami. (2014) \emph{Complex systems approximate matching approach for large graphs classification optimized by NSGA-II}, 2014 6th International Conference of Soft Computing and Pattern Recognition (SoCPaR), Tunis, 2014, pp. 112-117.
	
	\bibitem{mm1} M. Méndez. (2015) \emph{Set Operads in Combinatorics and Computer Science}. Springer Briefs in Mathematics. Springer.
	
	\bibitem{conf1}M. Méndez ET. AL. (2018) Hierarchical modeling of graphs using modular decomposition. Conference Abstract: 2nd International Neuroergonomics Conference.
	
	\bibitem{bnet}D. Meunier ET. AL. (2010) Modular and hierarchically modular organization of brain networks. Frontiers in Science. December 2010, Volume 4, Article 200,pp. 1--11.
	
	\bibitem{mr1}R.H. M\"ohring \& F.J. Rademacher. (1984) Substitution decomposition for discrete structures and connections with combinatorial optimization. \emph{Annals of Discrete Mathematics} 19, pp. 257--356.
	
	\bibitem{mh1}R.H. M\"ohring. (1985) Algorithmic aspects of the substitution decomposition in optimization over relations, set systems and boolean functions. \emph{Annals of Operations Research} 4(1985/6) pp. 195--225.
	\bibitem{palla} G. Palla ET. AL. (2010)  Multifractal network generator
	\emph{PNAS } April 27, 2010. 107 (17) 7640-7645.
	\bibitem{graphdrawing} C. Papadopoulos  \& C. Voglis. (2006) \emph{Drawing Graphs Using Modular Decomposition}. In: Healy P., Nikolov N.S. (eds) Graph Drawing. GD 2005. Lecture Notes in Computer Science, vol 3843. Springer, Berlin, Heidelberg.
		
		\bibitem{ros} R.A. Rossi \& N.K. Ahmed. (2015) The Network Data Repository with Interactive Graph Analytics and Visualization. \emph{Proceedings of the Twenty-Ninth AAAI Conference on Artificial Intelligence.} pp. 4292-4293.
	
	\bibitem{hergms} M. Schweinberger \&  M.S. Handcock. (2015) Local dependence in random graphs: characterization, properties, and statistical inference. \emph{Journal of the Royal Statistical Society: Series B (Statistical Methodology)}, Volume77, Issue3, pp. 647-676.
	
			\bibitem{size} A.E. Sizemore ET. AL. (2018) Cliques and cavities in the human connectome. \emph{J Comput Neurosci} Volume 44, Issue 1, pp. 115--145.
			
	\bibitem{tedder1}M. Tedder ET. AL. (2008)  \emph{Simpler Linear-Time Modular Decomposition Via Recursive Factorizing Permutations.} In: Aceto L., Damgård I., Goldberg L.A., Halldórsson M.M., Ingólfsdóttir A., Walukiewicz I. (eds) Automata, Languages and Programming. ICALP 2008. Lecture Notes in Computer Science, vol 5125. Springer, Berlin, Heidelberg.
	
\bibitem{wasfaust} Wasserman, S., \& Faust, K. (1994). Social network analysis: Methods and applications (Vol. 8). Cambridge University press.

\bibitem{zac} Zachary, W. (1977). An Information Flow Model for Conflict and Fission in Small Group., J. Anthro. Research 33(4), 452-473

\end{thebibliography}
\end{document}